\documentclass{article}

\usepackage{arxiv}
\usepackage{moreverb,url}
\usepackage{multirow}
\usepackage{amsmath}
\usepackage[colorlinks,bookmarksopen,bookmarksnumbered,citecolor=red,urlcolor=red]{hyperref}
\usepackage[utf8]{inputenc} % allow utf-8 input
\usepackage[T1]{fontenc}    % use 8-bit T1 fonts
\usepackage{hyperref}       % hyperlinks
\usepackage{url}            % simple URL typesetting
\usepackage{booktabs}       % professional-quality tables
\usepackage{amsfonts}       % blackboard math symbols
\usepackage{nicefrac}       % compact symbols for 1/2, etc.
\usepackage{microtype}      % microtypography
\usepackage{lipsum}
\usepackage{graphicx}
\title{Estimated Covid-19 burden in Spain: ARCH underreported non-stationary time series}

\author{
  David Mori\~na \\
  Department of Econometrics, Statistics and Applied Economics, Riskcenter-IREA\\ 
  Universitat de Barcelona\\ 
  Barcelona, Spain
  \texttt{dmorina@ub.edu} \\
   \And
   Amanda Fern\'andez-Fontelo \\
   Barcelona Graduate School of Mathematics (BGSMath) - Departament de Matem\`atiques\\ Universitat Aut\`onoma de Barcelona (UAB) \\
   Bellaterra, Spain \\
  \And
  Alejandra Caba\~na \\
  Barcelona Graduate School of Mathematics (BGSMath) - Departament de Matem\`atiques\\ Universitat Aut\`onoma de Barcelona (UAB) \\
  \And
  Argimiro Arratia \\
  Department of Computer Science\\
  Universitat Polit\`ecnica de Catalunya (UPC)\\
  Barcelona, Spain \\
  \And
  Pedro Puig \\
  Barcelona Graduate School of Mathematics (BGSMath) - Departament de Matem\`atiques\\ Universitat Aut\`onoma de Barcelona (UAB) \\
   Bellaterra, Spain \\
}

\begin{document}
\maketitle
\begin{abstract}
  The problem of dealing with misreported data is very common in a wide range of contexts for different reasons. The current situation caused by the Covid-19 worldwide pandemic is a clear example, where the data provided by official sources were not always reliable due to data collection issues and to the high proportion of asymptomatic cases. In this work, we explore the performance of Bayesian Synthetic Likelihood to estimate the parameters of a model capable of dealing with misreported information and to reconstruct the most likely evolution of the phenomenon. The performance of the proposed methodology is evaluated through a comprehensive simulation study and illustrated by reconstructing the weekly Covid-19 incidence in each Spanish Autonomous Community in 2020.
\end{abstract}

% keywords can be removed

\section{Introduction}\label{intro}
The Covid-19 pandemic that is hitting the world since late 2019 has made evident that having quality data is essential in the decision making chain, especially in epidemiology but also in many other fields. There is an enormous global concern around this disease, leading the World Health Organization (WHO) to declare public health emergency \cite{Sohrabi2020}. Many methodological efforts have been made to deal with misreported Covid-19 data, following ideas introduced in the literature since the late nineties \cite{Bernard2014,Arendt2013,Rosenman2006,Alfonso2015,Winkelmann1996,Gibbons2014}. These proposals range from the usage of multiplication factors \cite{Stocks2018} to Markov-based models \cite{Azmon2014,Magal2018} or spatio-temporal models \cite{Stoner2019}. Additionally, a new R \cite{RCoreTeam2019} package able to fitting endemic-epidemic models based on approximative maximum likelihood to underreported count data has been recently published \cite{JohannesBracher2019}. However, as a large proportion of the cases run asymptomatically \cite{Oran2020} and mild symptoms could have been easily confused with those of similar diseases at the beginning of the pandemic, its reasonable to expect that Covid-19 incidence has been notably underreported. Very recently several approaches based on discrete time series have been proposed \cite{Fernandez-Fontelo2016,FernandezFontelo2019,Fernandez-Fontelo2020} although there is a lack of continuous time series models capable of dealing with misreporting, a characteristic of the Covid-19 data and typically present in infectious diseases modeling. In this sense, a new approach for longitudinal data not accounting for temporal correlations is introduced in \cite{Morina2021} and a model capable of dealing with temporal structures using a different approach is presented in \cite{Morina2020}. A typical limitation of these kinds of models is the computational effort needed in order to properly estimate the parameters.

Synthetic likelihood is a recent and very powerful alternative for parameter estimation in a simulation based schema when the likelihood is intractable and, conversely, the generation of new observations given the values of the parameters is feasible. The method was introduced in \cite{Wood2010} and placed into a Bayesian framework in \cite{Price2018}, showing that it could be scaled to high dimensional problems and can be adapted in an easier way than other alternatives like approximate Bayesian computation (ABC). The method takes a vector summary statistic informative about the parameters and assumes it is multivariate normal, estimating the unknown mean and covariance matrix by simulation to obtain an approximate likelihood function of the multivariate normal.  
%%%%%%%%%%%%%%%%%%%%%%%%%%%%%%%%%%%%%%%%%%
\section{Materials and Methods}\label{methods}
AutoRegressive Conditional Heteroskedasticity (ARCH) models are a well-known approach to fitting time series data where the variance error is believed to be serially correlated. Consider an unobservable process $X_t$ following an AutoRegressive ($AR(1)$) model with ARCH(1) errors structure, defined by
$$
X_t = \phi_0 + \phi_1 \cdot X_{t-1} + Z_t,
$$

where $Z^2_t=\alpha_0+\alpha_1 \cdot Z^2_{t-1} + \epsilon_t,$ being $\epsilon_t \sim N(\mu_{\epsilon}(t), \sigma_{\epsilon}^2)$. The process $X_t$ represents the actual Covid-19 incidence. In our setting, this process $X_t$ cannot be directly observed, and all we can see is a part of it, expressed as

\begin{equation}\label{morina:eq1}
    Y_t=\left\{
                \begin{array}{ll}
                  X_t \text{ with probability } 1-\omega \\
                  q \cdot X_t \text{ with probability } \omega,
                \end{array}
              \right.
\end{equation}

where $q$ is the overall intensity of misreporting (if $0 < q < 1$ the observed process $Y_t$ would be underreported while if $q > 1$ the observed process $Y_t$ would be overreported) and $\omega$ can be interpreted as the overall frequency of misreporting (proportion of misreported observations). To model consistently the spread of the disease, the expectation of the innovations $\epsilon_t$ is linked to a simplified version of the well-known compartimental Susceptible-Infected-Recovered (SIR) model. At any time $t \in \mathbb{R}$ there are three kinds of individuals: Healthy individuals susceptible to be infected ($S(t)$), infected individuals who are transmitting the disease at a certain speed ($I(t)$) and individuals who have suffered the disease, recovered and cannot be infected again ($R(t)$). As shown in \cite{Fernandez-Fontelo2020}, the number of affected individuals at time $t$, $A(t) = I(t) + R(t)$ can be approximated by

\begin{equation}\label{eq:SIR}
  A(t) = \frac{M^{*}(\beta_0, \beta_1, \beta_2, t) A_0 e^{kt}}{M^{*}(\beta_0, \beta_1, \beta_2, t)+A_0(e^{kt}-1)},
\end{equation}
where $M^{*}(\beta_0, \beta_1, \beta_2, t) = \beta_0+\beta_1 \cdot C_1(t) + \beta_2 \cdot C_2(t)$, being $C_1(t)$ and $C_2(t)$ dummy variables indicating if time $t$ corresponds to a period where a mandatory confinment was implemented by the government and if the number of people with at least one dose of a Covid-19 vaccine in Spain was over 50\% respectively. At any time $t$ the condition $S(t) + I(t) + R(t) = N$ is fulfilled. The expression~(\ref{eq:SIR}) allow us to incorporate the behaviour of the epidemics in a realistic way, defining $\mu_{\epsilon}(t) = A(t) - A(t-1)$, the new affected cases produced at time $t$.

The Bayesian Synthetic Likelihood (BSL) simulations are based on the described model and the chosen summary statistics are the mean, standard deviation and the three first coefficients of autocorrelation of the observed process. Parameter estimation was carried out by means of the \textit{BSL} \cite{BSLManual,An2019} package for R \cite{RCoreTeam2019}. Taking into account the posterior distribution of the estimated parameters, the most likely unobserved process is reconstructed, resulting in a probability distribution at each time point. The prior of each parameter is set to be uniform on the corresponding feasible region of the parameter space and zero elsewhere.

\section{Results}
\label{results}
The performance and an application of the proposed methodology are studied through a comprehensive simulation study and a real dataset on Covid-19 incidence in Spain on this Section.

\subsection{Simulation study}\label{sim}
A thorough simulation study has been conducted to ensure that the model behaves as expected, including $ARCH(1)$, $AR(1)$, $MA(1)$ and $ARMA(1, 1)$ structures for the hidden process $X_t$ defined as

\begin{equation}\begin{array}{c}
  X_t = \phi_0 + \phi_1 \cdot X_{t-1} + Z_t, Z^2_t=\alpha_0+\alpha_1 \cdot Z^2_{t-1} + \epsilon_t \text{ (ARCH(1))} \\
  X_t = \phi_0 + \alpha \cdot X_{t-1} + \epsilon_t \text{ (AR(1))} \\
  X_t = \phi_0 + \theta \cdot \epsilon_{t-1} + \epsilon_t \text{ (MA(1))} \\
  X_t = \phi_0 + \alpha \cdot X_{t-1} + \theta \cdot \epsilon_{t-1} + \epsilon_t \text{ (ARMA(1, 1))}
\end{array}\end{equation}
where $\epsilon_t \sim N(\mu_{\epsilon}(t), \sigma_{\epsilon}^2)$.

The values for the parameters $\phi_1$, $\alpha_0$, $\alpha_1$, $\alpha$, $\theta$, $q$ and $\omega$ ranged from 0.1 to 0.9 for each parameter. Average absolute bias, average interval length (AIL) and average 95\% credibility interval coverage are shown in Table~\ref{tab:estim_sim}. To summarise model robustness, these values are averaged over all combinations of parameters, considering their prior distribution is a Dirac's delta with all probability concentrated in the corresponding parameter value. 

For each autocorrelation structure and parameters combination, a random sample of size $n = 1000$ has been generated using the R function \textit{arima.sim}, and the parameters $m=log(M^*)$ and $\beta$ have been fixed to $0.2$ and $0.4$ respectively. Several values for these parameters were considered but no substantial differences in the model performance were observed related to the value of these parameters or sample size, besides a poorer coverage for lower sample sizes, as could be expected.

\begin{table}[h]
\tiny\sf\centering
\caption{Model performance measures (average absolute bias, average interval length and average coverage) summary based on a simulation study.\label{tab:estim_sim}}
\begin{tabular}{ccccc}
\toprule
Structure & Parameter & Bias & AIL & Coverage (\%)\\
\midrule
\multirow{9}{*}{$ARCH(1)$}    & $\hat{\phi_0}$            &-0.377 & 3.586 & 68.77\% \\
                              & $\hat{\phi_1}$            & 0.122 & 0.525 & 66.08\% \\
                              & $\hat{\alpha_0}$          &-0.296 & 1.351 & 74.72\% \\
                              & $\hat{\alpha_1}$          &-0.085 & 0.920 & 77.34\% \\
                              & $\hat{\omega}$            &-0.020 & 0.234 & 83.71\% \\
                              & $\hat{q}$                 &-0.022 & 0.167 & 85.06\% \\
                              & $\hat{m}$                 &-0.226 & 0.783 & 79.17\% \\
                              & $\hat{\beta}$             &-0.734 & 3.581 & 76.83\% \\
                              & $\hat{\sigma_{\epsilon}}$ &-1.540 & 3.323 & 63.65\% \\
\hline
\multirow{7}{*}{$AR(1)$}      & $\hat{\phi_0}$            &-0.983 & 5.189 & 70.10\% \\
                              & $\hat{\alpha}$            & 0.043 & 0.814 & 92.46\% \\
                              & $\hat{\omega}$            &-0.003 & 0.111 & 94.10\% \\
                              & $\hat{q}$                 &-0.001 & 0.014 & 89.03\% \\
                              & $\hat{m}$                 & 0.001 & 0.190 & 75.17\% \\
                              & $\hat{\beta}$             & 0.007 & 0.192 & 74.49\% \\
                              & $\hat{\sigma_{\epsilon}}$ &-1.689 & 4.718 & 81.07\% \\
\hline
\multirow{7}{*}{$MA(1)$}      & $\hat{\phi_0}$            &-1.241 & 5.171 & 68.31\% \\
                              & $\hat{\theta}$            & 0.051 & 0.818 & 90.40\% \\
                              & $\hat{\omega}$            &-0.005 & 0.108 & 95.06\% \\
                              & $\hat{q}$                 &-0.001 & 0.014 & 87.24\% \\
                              & $\hat{m}$                 &-0.002 & 0.187 & 76.95\% \\
                              & $\hat{\beta}$             & 0.004 & 0.190 & 80.38\% \\
                              & $\hat{\sigma_{\epsilon}}$ &-1.619 & 4.679 & 83.95\% \\
\hline
\multirow{8}{*}{$ARMA(1, 1)$} & $\hat{\phi_0}$            & -1.834 & 5.107 & 61.01\% \\
                              & $\hat{\alpha}$            & 0.062  & 0.799 & 89.39\% \\
                              & $\hat{\theta}$            & 0.011  & 0.873 & 96.86\% \\
                              & $\hat{\omega}$            & -0.001 & 0.014 & 88.32\% \\
                              & $\hat{q}$                 & -0.005 & 0.109 & 94.97\% \\
                              & $\hat{m}$                 & 0.002  & 0.184 & 78.49\% \\
                              & $\hat{\beta}$             & 0.004  & 0.183 & 78.01\% \\
                              & $\hat{\sigma_{\epsilon}}$ & -1.828 & 4.631 & 74.74\% \\
\bottomrule
\end{tabular}
\end{table}

\subsection{Real incidence of Covid-19 in Spain}\label{covid}
\begin{figure*}[h]
\setlength{\fboxsep}{0pt}%
\setlength{\fboxrule}{0pt}%
\begin{center}
  \includegraphics[height=10cm, width=17cm]{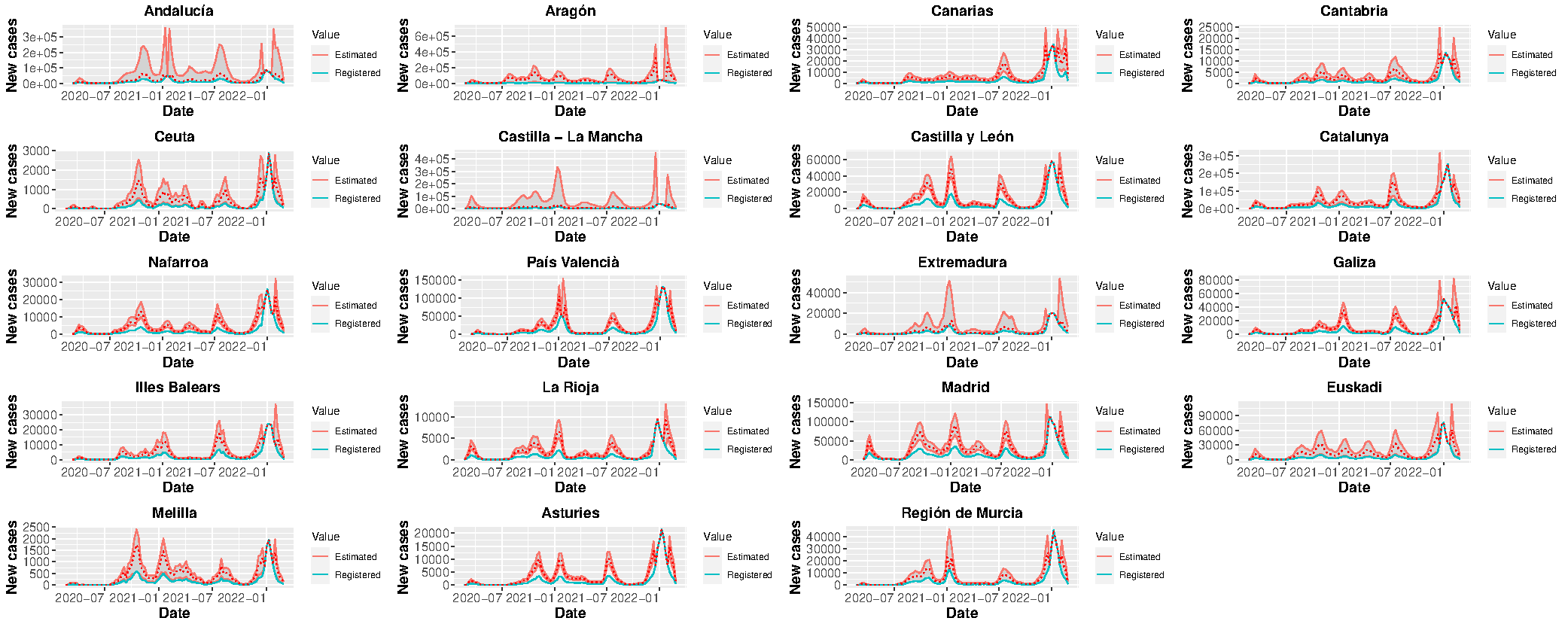}
  \caption{\label{morina:fig1} Registered and estimated weekly new Covid-19 cases in each Spanish region.}
  \end{center}
\end{figure*}

The betacoronavirus SARS-CoV-2 has been identified as the causative agent of an unprecedented world-wide outbreak of pneumonia starting in December 2019 in the city of Wuhan (China) \cite{Sohrabi2020}, named as Covid-19. Considering that many cases run without developing symptoms or just with very mild symptoms, it is reasonable to assume that the incidence of this disease has been underregistered. This work focuses on the weekly Covid-19 incidence registered in Spain in the period (2020/02/23-2022/02/27). It can be seen in Figure~\ref{morina:fig1} that the registered data (turquoise) reflect only a fraction of the actual incidence (red). The grey area corresponds to 95\% probability of the posterior distribution of the weekly number of new cases (the lower and upper limits of this area represent the percentile 2.5\% and 97.5\% respectively), and the dotted red line corresponds to its median.

\begin{figure*}[h]
\setlength{\fboxsep}{0pt}%
\setlength{\fboxrule}{0pt}%
\begin{center}
  \includegraphics[width=12cm]{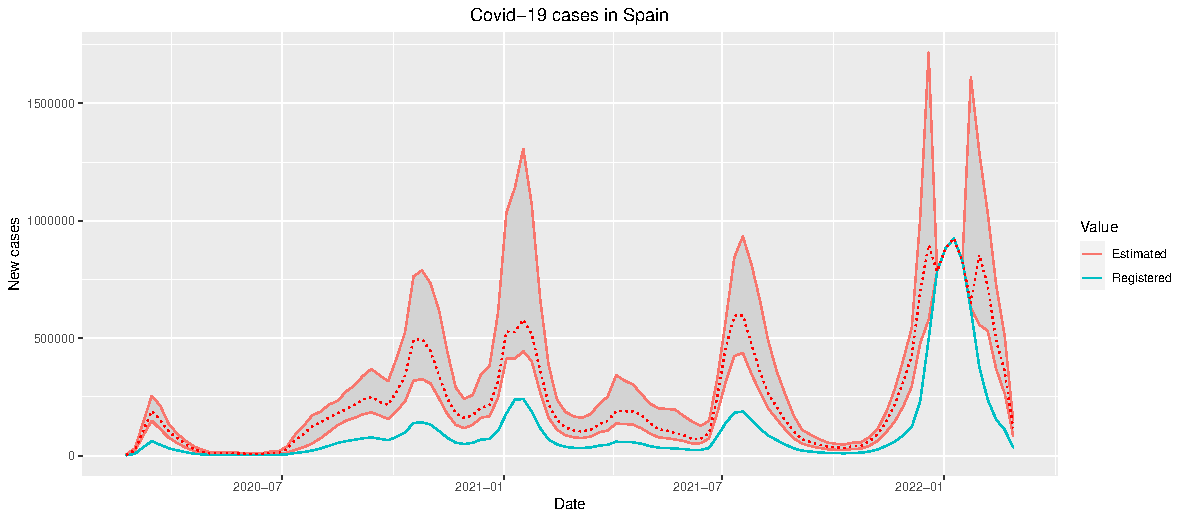}
  \caption{\label{morina:fig2} Registered and estimated weekly new Covid-19 cases globally in Spain.}
  \end{center}
\end{figure*}

In the considered period, the official sources reported 11,056,797 Covid-19 cases in Spain, while the model estimates a total of 25,283,406 cases (only 43.73\% of actual cases were reported). This work also revealed that while the frequency of underreporting is extremely high for all regions (values of $\hat{\omega}$ over 0.90 in all cases), the intensity of this underreporting is not uniform across the considered regions: Arag\'on is the CCAA with highest underreporting intensity ($\hat{q}=0.05$) while Extremadura is the region where the estimated values are closest to the number of reported cases ($\hat{q}=0.50$). Detailed underreported parameter estimates for each region can be found in Table~\ref{tab:ccaa_est}. Although the main impact of the vaccination programmes can be seen in mortality data, the results of this work also showed a significant decrease in the weekly number of cases as well in all CCAA except Arag\'on. Figure~\ref{morina:fig2} represents the estimated and registered Covid-19 weekly incidence globally for Spain. The estimated impact of the considered covariates is available in Appendix A (Table~\ref{tab:ccaa_est2}).

\begin{table}[h]
\small\sf\centering
\caption{Estimated underreported frequency and intensity for each Spanish CCAA. Reported values correspond to the median and percentiles 2.5\% and 97.5\% of the corresponding posterior distribution.\label{tab:ccaa_est}}
\begin{tabular}{ccc}
\toprule
CCAA & Parameter & Estimate (95\% CI)\\
\midrule
\multirow{2}{*}{Andaluc\'ia}  & $\hat{\omega}$  & 0.97 (0.95 - 0.99) \\
                              & $\hat{q}$       & 0.44 (0.41 - 0.48) \\
\midrule
\multirow{2}{*}{Arag\'on}    & $\hat{\omega}$  & 0.98 (0.97 - 0.99) \\
                             & $\hat{q}$       & 0.28 (0.27 - 0.32) \\
\midrule
\multirow{2}{*}{Asturies}    & $\hat{\omega}$  & 0.97 (0.90 - 0.99) \\
                             & $\hat{q}$       & 0.40 (0.37 - 0.53) \\
\midrule
\multirow{2}{*}{Cantabria}    & $\hat{\omega}$  & 0.97 (0.95 - 0.99) \\
                              & $\hat{q}$       & 0.30 (0.28 - 0.35) \\
\midrule
\multirow{2}{*}{Castilla y Le\'on}    & $\hat{\omega}$  & 0.98 (0.95 - 0.99) \\
                                      & $\hat{q}$       & 0.36 (0.32 - 0.41) \\
\midrule
\multirow{2}{*}{Castilla - La Mancha}    & $\hat{\omega}$  & 0.98 (0.96 - 0.99) \\
                                         & $\hat{q}$       & 0.33 (0.31 - 0.36) \\
\midrule
\multirow{2}{*}{Canarias}    & $\hat{\omega}$  & 0.98 (0.96 - 0.99) \\
                             & $\hat{q}$       & 0.35 (0.32 - 0.38) \\
\midrule
\multirow{2}{*}{Catalunya}    & $\hat{\omega}$  & 0.98 (0.96 - 0.99) \\
                              & $\hat{q}$       & 0.30 (0.27 - 0.34) \\
\midrule
\multirow{2}{*}{Ceuta}    & $\hat{\omega}$  & 0.98 (0.95 - 0.99) \\
                          & $\hat{q}$       & 0.28 (0.25 - 0.31) \\
\midrule
\multirow{2}{*}{Extremadura}    & $\hat{\omega}$  & 0.98 (0.95 - 1.00) \\
                                & $\hat{q}$       & 0.40 (0.36 - 0.44) \\
\midrule
\multirow{2}{*}{Galiza}     & $\hat{\omega}$  & 0.84 (0.33 - 0.98) \\
                            & $\hat{q}$       & 0.41 (0.35 - 0.56) \\
\midrule
\multirow{2}{*}{Illes Balears}    & $\hat{\omega}$  & 0.98 (0.96 - 0.99) \\
                                  & $\hat{q}$       & 0.36 (0.33 - 0.39) \\
\midrule
\multirow{2}{*}{Regi\'on de Murcia}    & $\hat{\omega}$  & 0.93 (0.45 - 0.98) \\
                                       & $\hat{q}$       & 0.46 (0.34 - 0.80) \\
\midrule
\multirow{2}{*}{Madrid}      & $\hat{\omega}$  & 0.98 (0.96 - 0.99) \\
                             & $\hat{q}$       & 0.37 (0.34 - 0.40) \\
\midrule
\multirow{2}{*}{Nafarroa}    & $\hat{\omega}$  & 0.99 (0.97 - 1.00) \\
                             & $\hat{q}$       & 0.30 (0.26 - 0.32)\\
\midrule
\multirow{2}{*}{Euskadi}    & $\hat{\omega}$  & 0.99 (0.97 - 0.99) \\
                            & $\hat{q}$       & 0.27 (0.25 - 0.31) \\
\midrule
\multirow{2}{*}{La Rioja}    & $\hat{\omega}$  & 0.98 (0.96 - 0.99) \\
                             & $\hat{q}$       & 0.31 (0.28 - 0.35) \\
\midrule
\multirow{2}{*}{Melilla}    & $\hat{\omega}$  & 0.97 (0.95 - 0.99) \\
                            & $\hat{q}$       & 0.34 (0.31 - 0.37) \\
\midrule
\multirow{2}{*}{Pa\'is Valenci\`a}    & $\hat{\omega}$  & 0.95 (0.40 - 0.98) \\
                                      & $\hat{q}$       & 0.46 (0.40 - 0.67) \\
\bottomrule
\end{tabular}
\end{table}

Figure~\ref{morina:fig2} shows the evolution of the registered (turquoise) and estimated (red) weekly number of Covid-19 cases in Spain in the period 2020/02/23-2022/02/27.

\begin{figure*}[h]
  \setlength{\fboxsep}{0pt}%
  \setlength{\fboxrule}{0pt}%
  \begin{center}
    \includegraphics[width=12cm]{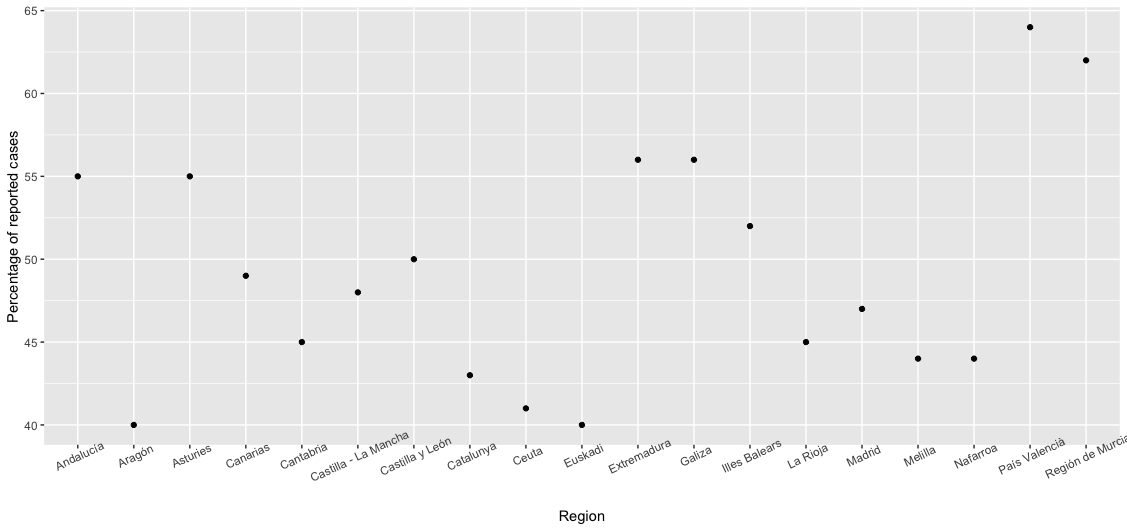}
    \caption{\label{morina:fig3} Percentage of reported cases in each CCAA.}
    \end{center}
  \end{figure*}

The global differences in underreporting magnitudes across regions can be represented by the percentage of reported cases in each CCAA (compared to model estimates), as shown in Figure~\ref{morina:fig3}.

\section{Discussion}\label{discussion}
Although it is very common in biomedical and epidemiological research to get data from disease registries, there is a concern about their reliability, and there have been some recent efforts to standardize the protocols in order to improve the accuracy of health information registries (see for instance \cite{Kodra2018,Harkener2019}). However, as the Covid-19 pandemic situation has made evident, it is not always possible to implement these recommendations in a proper way.  

The analysis of the Spanish Covid-19 data shows that in average only around 60\% of the cases in the period 2020/02/23-2022/02/27 were reported, and that there are significant differences in the severity of underreporting across the regions. The impact of the vaccination programme can also be assessed, achieving a significant decrease in the Covid-19 incidence in almost all regions after 50\% of the population had one dose at least (although these results would probably be notably different if including SARS-CoV-2 immunity-escape variants like BA.4 or BA.5, which are currently predominant in many countries), while the impact of the mandatory lockdown could only be detected by the model in 7 regions. 

Having accurate data is key in order to provide public health decision-makers with reliable information, which can also be used to improve the accuracy of dynamic models aimed to estimate the spread of the disease \cite{Zhao2020} and to predict its behavior. 

The proposed methodology can deal with misreported (over- or under-reported) data in a very natural and straightforward way, and is able to reconstruct the most likely hidden process, providing public health decision-makers with a valuable tool in order to predict the evolution of the disease under different scenarios. 

The simulation study shows that the proposed methodology behaves as expected and that the parameters used in the simulations, under different autocorrelation structures, can be recovered, even with severely underreported data.

\section*{Acknowledgements}
Investigation funded by Fundaci\'on MAPFRE. This work was partially supported by grant RTI2018-096072-B-I00 from the Spanish Ministry of Science and Innovation.

\bibliographystyle{unsrt}  


  \section*{Appendix A}

\begin{table}[h]
\small\sf\centering
\caption{Impact of covariates for each Spanish CCAA. Reported values correspond to the median and percentiles 2.5\% and 97.5\% of the corresponding posterior distribution.\label{tab:ccaa_est2}}
\begin{tabular}{ccc}
\toprule
CCAA & Parameter & Estimate (95\% CI)\\
\midrule
\multirow{2}{*}{Andaluc\'ia}  & $\hat{Vacc}$  & -1.71 (-2.66, -0.68) \\
                              & $\hat{Conf}$  & -1.67 (-2.31, -0.39) \\
\midrule
\multirow{2}{*}{Arag\'on}    & $\hat{Vacc}$  & -1.06 (-1.36, -0.69) \\
                             & $\hat{Conf}$  & 0.76 (0.17, 1.43) \\
\midrule
\multirow{2}{*}{Asturies}    & $\hat{Vacc}$  & -0.90 (-1.77, -0.63) \\
                             & $\hat{Conf}$  & 0.44 (0.11, 0.69) \\
\midrule
\multirow{2}{*}{Cantabria}    & $\hat{Vacc}$  & -0.53 (-1.29, -0.25) \\
                              & $\hat{Conf}$  & -0.44 (-0.71, 0.002) \\
\midrule
\multirow{2}{*}{Castilla y Le\'on}    & $\hat{Vacc}$  & -1.22 (-1.88, -0.60) \\
                                      & $\hat{Conf}$  & -0.84 (-1.33, -0.23) \\
\midrule
\multirow{2}{*}{Castilla - La Mancha}    & $\hat{Vacc}$  & -0.80 (-1.11, -0.40) \\
                                         & $\hat{Conf}$  & 0.06 (-0.18, 0.44) \\
\midrule
\multirow{2}{*}{Canarias}    & $\hat{Vacc}$  & -1.34 (-1.78, -1.06) \\
                              & $\hat{Conf}$ & -0.92 (-2.06, -0.29) \\
\midrule
\multirow{2}{*}{Catalunya}    & $\hat{Vacc}$  & -1.51 (-1.97, -0.94) \\
                              & $\hat{Conf}$  & -0.25 (-0.52, 0.21) \\
\midrule
\multirow{2}{*}{Ceuta}    & $\hat{Vacc}$  & -1.38 (-1.93, -0.84) \\
                          & $\hat{Conf}$  & 0.007 (-0.52, 0.34) \\
\midrule
\multirow{2}{*}{Extremadura}    & $\hat{Vacc}$  & -0.72 (-1.30, -0.37) \\
                                & $\hat{Conf}$  & 1.45 (1.24, 1.83) \\
\midrule
\multirow{2}{*}{Galiza}     & $\hat{Vacc}$  & -2.03 (-3.07, -1.34) \\
                            & $\hat{Conf}$  & -0.20 (-0.53, 0.18) \\
\midrule
\multirow{2}{*}{Illes Balears}    & $\hat{Vacc}$  & -0.72 (-1.16, -0.34) \\
                                  & $\hat{Conf}$  & 0.74 (0.43, 1.01) \\
\midrule
\multirow{2}{*}{Regi\'on de Murcia}    & $\hat{Vacc}$  & -1.97 (-3.07, -0.59) \\
                                       & $\hat{Conf}$  & 0.62 (-0.02, 1.36) \\
\midrule
\multirow{2}{*}{Madrid}      & $\hat{Vacc}$  & -0.35 (-0.77, -0.07) \\
                             & $\hat{Conf}$  & 0.35 (-0.39, 0.59) \\
\midrule
\multirow{2}{*}{Nafarroa}   & $\hat{Vacc}$  & -2.05 (-3.20, -1.33) \\
                            & $\hat{Conf}$  & -1.71 (-1.92, -0.53) \\
\midrule
\multirow{2}{*}{Euskadi}    & $\hat{Vacc}$  & -0.10 (-0.24, 0.00) \\
                            & $\hat{Conf}$  & -0.42 (-0.69, -0.21) \\
\midrule
\multirow{2}{*}{La Rioja}    & $\hat{Vacc}$  & -0.43 (-0.71, -0.22) \\
                             & $\hat{Conf}$  & -0.83 (-1.08, -0.35) \\
\midrule
\multirow{2}{*}{Melilla}   & $\hat{Vacc}$  & -1.59 (-2.05, -0.93) \\
                           & $\hat{Conf}$  & -0.48 (-0.82, -0.11) \\
\midrule
\multirow{2}{*}{Pa\'is Valenci\`a}    & $\hat{Vacc}$  & -1.70 (-2.64, -0.52) \\
                                      & $\hat{Conf}$  & 1.45 (1.24, 1.83) \\
\bottomrule
\end{tabular}
\end{table}

\end{document}